\documentclass[11pt]{article}
\usepackage{amsmath}
\usepackage{amsfonts}
\usepackage{amssymb}
\usepackage{graphicx}

\def\bea{\begin{eqnarray}}
\def\eea{\end{eqnarray}}

\begin{document}
\begin{center}
\LARGE { \bf Correlation Functions for Galilean Conformal Invariant Field Theory
  }
\end{center}
\begin{center}
{\bf M. R. Setare\footnote{rezakord@ipm.ir} \\  V. Kamali\footnote{vkamali1362@gmail.com}}\\
 {\ Department of Science of Bijar, University of  Kurdistan  \\
Bijar, IRAN.}
 \\
 \end{center}
\vskip 3cm

\begin{abstract}
In this paper we consider non-relativistic-conformal group, then we calculate two point function for the fields that are Galilean conformal-invariant, then we show that the correlation function for Galilean conformal-invariant fields in $d=2$ are different from $d\neq2$. Our result in $d\neq2$ is different with previous results in \cite{3}.

\end{abstract}

\newpage

\section{Introduction}
Recently, there has been some interest in extending the AdS/CFT correspondence to non-relativistic field theories (\cite{1},\cite{2}). The study of a different non-relativistic limit was initiated in \cite{ba}, where the non-relativistic conformal symmetry obtained by a parametric contraction of the relativistic conformal group. The Galilean conformal group arises as a contraction relativistic conformal group, where in $d=3$ it is a fifteen parameter group which contains the ten parameter Galilean subgroup. Beside Galilean conformal algebra (GCA), there is another Galilean algebra, the twelve parameter schrodinger algebra. The dilatation generator in the schrodinger group scales space and time differently, $x_i\rightarrow \lambda x_i$, $t\rightarrow \lambda^2 t$, but in contrast the corresponding generator in GCA scales space and time in the same way, $x_i\rightarrow \lambda x_i$, $t\rightarrow \lambda t$. In this paper, we calculate the two point function in the Galilean conformal algebra separately in $d=2$ and $d\neq 2$ dimensions. We show that the form of two point function in $d=2$ dimension has a difference with another dimensions. Our result for $d=2$ is similar to the result of \cite{3}, but in contrast to this reference, our result for $d\neq 2$ has another form.\\ The paper organized as follows. Section 2 is a brief review of GCA. In section 3, 4  we show the representations of GCA in  $d=2$  and  $d\neq 2$  dimensions respectively. Then in section 5, we calculate the form of two point function in $d=2$ and $d\neq 2$ dimensions. Finally, in section 6, we close by some concluding remarks.
\\

\section{Galilean conformal algebras }
Galilean conformal  algebras (GCA) was obtained via a direct contraction of conformal generators. Physically,  this come from taking
$ t\rightarrow t~~~x_{i}\rightarrow \epsilon x_{i}~~~\epsilon\rightarrow 0 $.
 The generators of conformal group are

\begin{eqnarray}\label{1}
  P_{\mu}=-i\partial_{\mu},~~~~~D=-i x^{\mu}\partial_{\mu},~~~~~J_{\mu\nu}=i(x_{\mu}\partial_{\nu}-x_{\nu}\partial_{\mu}),~~~\\
  \nonumber
  K_{\mu}=-i(2x_{\mu}x^{\nu}\partial_{\nu}-x^2\partial_{\mu})~~~~~~~~~~~~~~~~~~~~~~~~~~~~~~~~~~~~
\end{eqnarray}

(where $\mu,\nu=0,1,...,d$).\\
 From the above scaling we obtain the Galilean conformal vector field generators

\begin{eqnarray}\label{2}
 P_{i}=\partial_{i}~~~~H=\partial_{t}~~~D=(t\partial_{t}+x^{i}\partial_{i})~~~J_{ij}=-(x_{i}\partial_{j}-x_{j}\partial_{i})~~~~B_{i}=-t\partial_{i}\\
 \nonumber
 K=K_0=(2tx^{i}\partial_{i}+t^2\partial_{t})~~~~~~~~~~~K_{i}=-t^2\partial_{i}~~~~~~~~~~~~~~~~~~~~~~~~~~~~~~~~~~~~~~
\end{eqnarray}

$P_{i},H,D,J_{ij},B_{i},K,K_{i}$ are  spatial translations, time translation, rotations, boosts, time component and space components of special conformal transformation respectively. These generators obey the following commutation relations, where define the Galilean conformal algebras.

\begin{eqnarray}\label{3}
  [J_{ij},J_{rs}]=So(d)~~~~~~~~~~~~~~~~~~~~~~~~[J_{ij},B_{r}]=-(B_{i}\delta_{ji}-B_{j}\delta_{ir})\\
  \nonumber
  [J_{ij},P_{r}]=-(P_{i}\delta_{ji}-P_{j}\delta_{ir})~~~~~~~~~~~~~~~~~~~~~~~~~~~~~~~~[J_{ij},H]=0\\
  \nonumber
  [H,B_{i}]=-P_{i}~~~~~~~~~~~~~~~~~~~~~~~~~~~~~~~~~~~~~~~~~~~~~~~~~[P_{i},B_{j}]=0\\
  \nonumber
  [D,H]=-H~~~~~~~~~~~~~~~~~~~~~~~~~~~~~~~~~~~~~~~~~~~~~~~[D,P_{i}]=-P_{i}\\
  \nonumber
  [D,J_{ij}]=0~~~~~~~~~~~~~~~~~~~~~~~~~~~~~~~~~~~~~~~~~~~~~~~~~~[K,P_{i}]=2B_{i}\\
  \nonumber
  [K,B_{i}]=-K_{i}~~~~~~~~~~~~~~~~~~~~~~~~~~~[J_{ij},K_{r}]=-(K_{i}\delta_{ji}-K_{j}\delta_{ir})\\
  \nonumber
  [D,K]=K~~~~~~~~~~~~~~~~~~~~~~~~~~~~~~~~~~~~~~~~~~~~~~~~[D,K_{i}]=-K_{i}\\
  \nonumber
  [K,H]=-2D~~~~~~~~~~~~~~~~~~~~~~~~~~~~~~~~~~~~~~~~~~~~~~[H,K_{i}]=2B_{i}
\end{eqnarray}

\section{ Representations of  Galilean conformal group in $d\neq2$ dimension }
We start by studying the subgroup of the galilean  group that leaves the point $x=0$ invariant that is non-relativistic-Lorentz group (with previous contraction). We then introduce  matrix representations $ S_{0i}$ and $S_{ij}$ to define the action of infinitesimal boost and infinitesimal rotation transformations on the field $\phi(0)$

\begin{equation}\label{4}
B_{i}\phi(0)=J_{0i}\phi(0)=S_{0i}\phi(0)~~~~~~~~~~~~~~~~~~~~J_{ij}\phi(0)=S_{ij}\phi(0)
\end{equation}

$S_{0i}$ and $S_{ij}$ are  spin operators associated with the field $\phi$,
next by use of the commutation relations of the GCA (\ref{3}), we translate the generator $J_{0i}$
to a nonzero value of $x$

\begin{equation}\label{5}
\exp -i(tH+x^{i}P_{i})J_{0i}\exp i(tH+x^{i}P_{i})=S_{0i}-tP_{i}
\end{equation}

\begin{equation}\label{6}
\exp -i(tH+x^{i}P_{i})J_{ij}\exp i(tH+x^{i}P_{i})=S_{ij}-(x_{i}P_{j}-x_{j}P_{i})
\end{equation}

The above translations are explicitly calculated by use of the Hausdorff formula.
 These  allow us to write the action of the generators

\begin{equation}\label{7}
P_{i}=\partial_{i}\phi~~~~~~~~~~~~~~~~~~~~H\phi(x)=\partial_{t}\phi
\end{equation}

\begin{equation}\label{8}
 B_{i}=J_{0i}=-t\partial_{i}\phi+S_{0i}\phi~~~~~~~~~~~~~~~J_{ij}=-(x_{i}\partial_{j}-x_{j}\partial_{i})\phi+S_{ij}\phi
\end{equation}

 We proceed in the same way for the full GCA group. The subgroup that leaves the origin $x=0$ invariant is generated by rotations and special transformation. If we remove the time translation generator, and space translation generators from the GCA, we then denote by $ S_{ij},S_{0i},\widetilde{\Delta},k$ and $,k_{i}$ the respective value of the generators $J_{ij},B_{i},D,K$ and $K_{i}$ at $x=0$.
 These  form a matrix representation of the reduced GCA.

\begin{eqnarray}\label{9}
[\widetilde{\Delta} ,S_{0i}]=0~~~~~~~~~~~~~~~~~~~~~~~~~~~~~~~~~~~~[\widetilde{\Delta }, S_{ij}]=0 \\
\nonumber
 [\widetilde{\Delta},k_{i}]=-k_{i}~~~~~~~~~~~~~~~~~~~~~~~~~~~~~~~~~~~~[\widetilde{\Delta}, k]=k \\
 \nonumber
  [S_{ij},k]=0~~~~~~~~~~~~~~~~~[S_{ij},k_{r}]=-(k_{i}\delta_{jr}-k_{j}\delta_{jr}) \\
  \nonumber
  [S_{0i},k]=-k_{i}~~~~~~~~~~~~~~~~~~~~~~~~~~~~~~~~~[S_{0i},k_{j}]=0 \\
  \nonumber
   [k_{i},k_{j}]=0~~~~~~~~~~~~~~~~~~~~~~~~~~~~~~~~~~~~~~~[k,k_{i}]=0\\
   \nonumber
   [S_{ij},S_{0r}]=-(S_{0i}\delta_{jr}-S_{0j}\delta_{jr})~~~~~~~~~~~~~~~~~~~~~~~~~~~
\end{eqnarray}

The commutation relations  (\ref{9})  allow us to translate the generators, using the Hausdorff formula

\begin{eqnarray}\label{10}
\exp i(tH+x^ip_i)D\exp-i(tH+x^ip_i)=D+tH+x^ip_i
\end{eqnarray}

\begin{eqnarray}\label{11}
\exp i(tH+x^ip_i)K_i\exp-i(tH+x^ip_i)=K_i+2tS_{0i}-\frac{t^2}{2}(2p_i)
\end{eqnarray}

\begin{eqnarray}\label{12}
\exp i(tH+x^ip_i)K\exp-i(tH+x^ip_i)=K+ 2t\widetilde{\Delta}+t^2H+2x^iS_{0i}+2tx^iP_i
\end{eqnarray}
 From the above relations we obtain the following extra transformation relations with semi-classic notation,

\begin{eqnarray}\label{13}
[D,\phi]=(x^i\partial_{i}+t\partial_{t}+\widetilde{\Delta})\phi~~~~~~~~~~~~~~~~~~~~~~~~~~~~~\\
\nonumber
 [K_{i},\phi]=(-t^2\partial_t +2tS_{0i})\phi~~~~~~~~~~~~~~~~~~~~~~~~~~~~~\\
 \nonumber
 [K,\phi]=(t^2\partial_t+2tx^i\partial_i+2t\widetilde{\Delta}-2x^iS_{0i})\phi~~~~~~~~~~~
\end{eqnarray}

 If we demand that the field $\phi(x)$ belong to an irreducible representation of the non-relativistic-Lorentz group
 then by Schur's lemma,  any matrix that commutes with all the generators should be a multiple of the identity. If $\widetilde{\Delta}$ is a constant then it commute with all generators of GCA
 and from commutation relations (\ref{9}) $S_{ij},S_{0i},k,k_{i}$ are equal to zero and finally we obtain following relations

\begin{eqnarray}\label{14}
[H,\phi]=\partial_t \phi~~~~~~~~~~~~~~~~~~~~~~~~~~~~~~~~~~~~~~~~~~~\\
\nonumber
[P_i,\phi]=\partial_{i}\phi~~~~~~~~~~~~~~~~~~~~~~~~~~~~~~~~~~~~~~~~~~~\\
\nonumber
[D,\phi]=(x^i\partial_i+t\partial_t+\Delta)\phi~~~~~~~~~~~~~~~~~~~~~~~~~\\
\nonumber
[K_i,\phi]=-t^2\partial_t \phi~~~~~~~~~~~~~~~~~~~~~~~~~~~~~~~~~~~~~~\\
\nonumber
[B_i,\phi]=-t\partial_t \phi~~~~~~~~~~~~~~~~~~~~~~~~~~~~~~~~~~~~~~~~\\
\nonumber
[K,\phi]=(t^2\partial_t+2tx^i\partial_i+2t\Delta)\phi~~~~~~~~~~~~~~~~~~~\\
\nonumber
 [J_{ij},\phi]=-(x_i\partial_{j}-x_j\partial_i)~~~~~~~~~~~~~~~~~~~~~~~~~~~~
\end{eqnarray}

From the above relations, one can obtain  the change in $\phi$ under a finite GCA transformation. Under a GCA transformation $x\rightarrow x'$, a spinless field $\phi(x)$ transforms as

\begin{eqnarray}\label{15}
\phi'(x')=|\frac{\partial x'}{\partial x}|^{-\frac{\Delta}{d}}\phi(x)
\end{eqnarray}

where $|\frac{\partial x'}{\partial x}|$ is the Jacobian of the GCA  coordinates transformations  and $\Delta$ is conformal dimension of $\phi$. A field
transforming like (\ref{15}) is a non-relativistic quasi-primary field.

\section{Representation of the Galilean conformal group in d=2 dimension}

In this section we consider the Galilean conformal group in d=2 where the rotation is absent, so we have following  transformation relations with semi-classic notation. (In this section  we introduce $S_{0x}=\xi $ which is the spin operator associated with the field $\phi$.)

\begin{eqnarray}\label{16}
[D,\phi]=(x\partial_x+t\partial_t+\Delta)\phi~~~~~~~~~~~~~~~~~~~~~~~~~~~~~~~~~~~~~~~~~~~~~\\
\nonumber
[K_x,\phi]=(-t^2\partial_t+2t \xi) \phi~~~~~~~~~~~~~~~~~~~~~~~~~~~~~~~~~~~~~~~~~~~~~~\\
\nonumber
[K,\phi]=(t^2\partial_t+2tx\partial_x+2(t\Delta-x\xi))\phi~~~~~~~~~~~~~~~~~~~~~~~~~~~~~\\
\nonumber
[B,\phi]=(-t\partial_{x}+\xi)~~~~~~~~~~~~~~~~~~~~~~~~~~~~~~~~~~~~~~~~~~~~~~~~~~~~~
\end{eqnarray}

  $\widetilde{\Delta}$ commutes with non-relativistic-Lorentz group generators then by Schur's lemma $\widetilde{\Delta}$ should be constant and $K=K_i=0$.
$\xi$ commutes with all generators of Galilean conformal-group so using Schur's lemma $\xi$ is a constant. Also in $d=2$ the definition of non-relativistic quasi-primary field applies  to fields with spin.

\section{Non-relativistic conformal correlation functions}

In this section we would like to find the form of  two point function for the GCA. According to the  above results, we have different representation of the Galilean conformal group in $d=2$ and $d\neq2$. Now we consider equations (\ref{14}) and (\ref{16}), and calculate the two point function of the GCA separately.

\subsection{Two point function in $d=2$}
We introduce  two point function as

\begin{eqnarray}\label{17}
F=<0|\phi_1(r_1,t_1)\phi_2(r_2,t_2)|0>
\end{eqnarray}

  where $\phi_1$ and $\phi_2$ are non-relativistic quasi-primary fields with conformal dimensions and spins   ($\Delta_1$,$\xi_1$) and ($\Delta_2$,$\xi_2$) respectively (in $d\neq2$  $\xi_1=\xi_2=0$). In two dimension  we demand space and time translations invariant on $F$, so $F$ depends on the $\tau=t_1-t_1$ and $r=x_1-x_2$. From previous section we get three  equations which admit  constrain on the form of two point function.
 $\phi_1$ and $\phi_{2}$ are Galilean boost invariant

\begin{eqnarray}\label{18}
<0\mid[B,\phi_1\phi_2]\mid0>=0  ~~~~~~~~~~~
\end{eqnarray}

so
\begin{eqnarray}\label{19}
 (-\tau\partial_{r}+\xi)F=0~~~~~~~~~~~
\end{eqnarray}

where $\xi=\xi_1+\xi_2$. The solution of (\ref{19}) can be written as

\begin{eqnarray}\label{20}
F=C(\tau)\exp(\frac{\xi r}{\tau})
\end{eqnarray}

 where $C(\tau)$ is an arbitrary function of $\tau$.  $\phi_1$ and $\phi_2$
 are dilatation invariant also

\begin{eqnarray}\label{21}
<0\mid[D,\phi_1\phi_2]\mid0>=0
\end{eqnarray}
so
\begin{eqnarray}\label{22}
(\tau\partial_{\tau}+r\partial_{r}+\Delta)F=0
\end{eqnarray}

the solution of (\ref{22}) can be written as

\begin{eqnarray}\label{23}
  F=C'\tau^{-\Delta}\exp(\frac{\xi r}{\tau})~~~~
\end{eqnarray}
where $C'$ is an arbitrary constant.
 We still have two another conditions

\begin{eqnarray}\label{24}
<0\mid[K_{x},\phi_1\phi_2]\mid0>=0,~~~~~~~~~~~~~<0\mid[K,\phi_1\phi_2]\mid0>=0
\end{eqnarray}
Using these conditions, we obtain $\Delta_1=\Delta_2$ and $\xi_1=\xi_2$.
 So, the final form of the two point function for Galilean conformal field theory in two dimension is as

\begin{eqnarray}\label{25}
F(r,\tau)=C' \delta_{\Delta_1,\Delta_2}\delta_{\xi_1,\xi_2}\tau^{-2\Delta_1}\exp(\frac{2\xi r}{\tau})
\end{eqnarray}

Two point function of GCA in two dimension also can be obtained via contraction method of \cite{4}.

\begin{eqnarray}\label{26}
<0\mid\phi_1(x_1,t_1)\phi_2(x_2,t_2)\mid0>_{GCA}~~~~~~~~~~~~~~~~~~~~~~~~~~~~~~~~~~~~~~~~~~~~\\
\nonumber
=\lim_{\epsilon\longrightarrow0}<0\mid\phi_1(x_1,t_1)\phi_2(x_2,t_2)\mid0>_{CFT}~~~~~~~~~~~~~~~~~~~~~~~~~~~~~~~~
\end{eqnarray}
so
\begin{eqnarray}\label{27}
<0\mid\phi_1(x_1,t_1)\phi_2(x_2,t_2)\mid0>_{GCA}=\lim_{\epsilon\longrightarrow0}\delta_{h_{1},h_{2}}\delta_{\overline{h}_1\overline{h}_2}
Z_{12}^{-2h_1}\overline{Z}_{12}^{-2\overline{h}_{1}}~~~~~~~~~~~~
\end{eqnarray}
where

\begin{eqnarray}\label{28}
Z_{12}=t_{12}+i\epsilon x_{12},  ~~~~~~~~~~~~~~~~~ \overline{Z}=t_{12}-i\epsilon x_{12}
\end{eqnarray}
\begin{eqnarray}\label{29}
h=\Delta+\frac{i\xi}{\epsilon},~~~ ~~~~~~~~~~~~~~~~~~~~~~~\overline{h}=\Delta -\frac{i\xi}{\epsilon}
\end{eqnarray}
Therefore
\begin{eqnarray}\label{30}
F=<0|\phi_1(r_1,t_1)\phi_2(r_2,t_2)|0>~~~~~~~~~~~~~~~~~~~~~~~~~~~~~~~~~~~~~~~~\\
\nonumber
=\lim_{\epsilon\longrightarrow0}\delta_{h_{1},h_{2}}\delta_{\overline{h}_1\overline{h}_2}(t_{12}+i\epsilon x_{12})^{-2(\Delta+\frac{i\xi}{\epsilon})}(t_{12}-i\epsilon x_{12})^{-2(\Delta -\frac{i\xi}{\epsilon})}
\end{eqnarray}
or in another form
\begin{eqnarray}\label{31}
F(r,t_{12})=C' \delta_{\Delta_1,\Delta_2}\delta_{\xi_1,\xi_2}t_{12}^{-2\Delta_1}\exp(\frac{2\xi r}{t_{12}})~~~~~~~~~~~~~~~~~~
\end{eqnarray}
\subsection{Two point function in $d\neq 2$}
We introduce two point function as

\begin{eqnarray}\label{32}
F(x_{1}^{i},t_1,x_{2}^{i},t_2)=<0|\phi_1(x_1^{i},t_1)\phi_2(x_2^{i},t_2)|0>
\end{eqnarray}
 where $\phi_1$  and $\phi_2$ are non-relativistic quasi-primary fields with conformal dimensions $\Delta_1$ and $\Delta_2$. We demand space and time translation invariant on the $F$, so $F$ depends on the $\tau=t_1-t_2$ and $r^{i}=x_{1}^{i}-x_{2}^{i}$. From previous section we have three  equations which admit some constrain on the form of two point function. From Galilean boosts invariant we have

\begin{eqnarray}\label{33}
<0\mid[B_i,\phi_1\phi_2]\mid0>=0  ~~~~~~~~~~~~~~~~~~~~~~~~~~~~~~~~~~~~~~~~~~
\end{eqnarray}
then
\begin{eqnarray}\label{34}
 (-\tau\partial_{r_{i}})F=0~~~~~~~~~~~~~~~~~~~~~~~~~~~~~~~~~~~~~~~~~~~~~~~~~
\end{eqnarray}
so
\begin{eqnarray}\label{35}
 F=F(\tau)~~~~~~~~~~~~~~~~~~~~~~~~~~~~~~~~~~~~~~~~~~~~~~~~~~~~~~~~~~~~
\end{eqnarray}

From dilatation invariant we obtain

\begin{eqnarray}\label{36}
<0\mid[D,\phi_1\phi_2]\mid0>=0  ~~~~~~~~~~~~~~~~~~~~~~~~~~~~~~~~~~~~~~~~
\end{eqnarray}
then
\begin{eqnarray}\label{37}
(\tau\partial_{\tau}+r^{i}\partial_{r^{i}}+\Delta)F=0~~~~~~~~~~~~~~~~~~~~~~~~~~~~~~~~~~~~~
\end{eqnarray}
therefore
\begin{eqnarray}\label{38}
  F=C\tau^{-\Delta}~~~~~~~~~~~~~~~~~~~~~~~~~~~~~~~~~~~~~~~~~~~~~~~~~~~~~~~~~~
\end{eqnarray}
where $C$ is an arbitrary constant.
From special conformal invariant,

\begin{eqnarray}\label{39}
<0\mid[K_{i},\phi_1\phi_2]\mid0>=0,~~~~~~~~~~~<0\mid[K,\phi_1\phi_2]\mid0>=0~
\end{eqnarray}
 we obtain the following constrain on the conformal dimensions

\begin{eqnarray}\label{40}
\Delta_1=\Delta_2
\end{eqnarray}

Finally the complete  form of the two point function in $d\neq2$ is

\begin{eqnarray}\label{41}
F(r,\tau)=C\delta_{\Delta_{1},\Delta_{2}}\tau^{-2\Delta_1}
\end{eqnarray}
 As we see, GCA completely specifies the form of the two-point function in any number of space
 dimensions,
 but these  results deserve some comments. It is instructive to compare them with
 the two-point function obtained from the requirement of Galilean conformal invariance in \cite{3}.
  Our result for correlation function in $d\neq2$ did not depend to the rapidity. The exponential behaviour
   of the two-point scaling function in (\ref{25}) is a consequence of boost invariance. If we had considered
    the GCA in $d\neq2$ the Boost generators are different \cite{3}, our result for $B_i$ in $d\neq2$ did not
     depend to the rapidity, this is due to the commutation relation
      $[S_{ij},S_{0r}]=-(S_{0i}\delta_{jr}-S_{0j}\delta_{jr})$ (we have this relation in $d\neq2$ only).
       \\From the above relation, one can obtain the $S_{0i}=0$, so the eigenvalue of these generators,
        are zero also, i.e $\xi_i=0$, in $d\neq2$. So the result of correlation  function in $d\neq2$, Eq.(\ref{41}) did not depend to the rapidity.\\

\section{Conclusion}
In this paper we have calculated the two point function of Galilean conformal invariant field theory in $d=2$ and $d\neq 2$ dimensions separately. Previously, the author of (\cite{3}, \cite{4}) have calculated the two point functions of GCA in all dimensions. The result of \cite{3} is similar to our result (\ref{31}), but the result of \cite{5} is similar to our result (\ref{41}). The authors of \cite{5} have considered  the states with zero rapidity. So the difference of there results with \cite{3} arises from this point. Here we have chosen the state with non-zero rapidity, but in contrast to \cite{3}, our result for correlations functions in $d\neq 2$ did not depend to the rapidity. This is due to the last commutation relation of (\ref{9}), i.e. $[S_{ij},S_{0r}]=-(S_{0i}\delta_{jr}-S_{0j}\delta_{jr})$ , we have this relation for $d\neq 2$ only, so it is naturally there was a difference between $d=2$ and $d\neq 2$. For the fields with zero eigenvalue of $S_{ij}$, from the mentioned commutation relation, one can obtain the $S_{0i}=0$, so the eigenvalue of these generators, which are rapidity are zero also, i.e. $\xi_i=0$. Therefore in $d\neq 2$ the two point functions have not depend to $\xi_i$, and the correlation function in these dimensions take the form of (\ref{41}).

\end{document}